\documentclass{emulateapj}\usepackage{natbib}\usepackage{times}

\begin{document}

\shorttitle{GRBs and the High-$z$ SFR}
\shortauthors{ Y{\"U}KSEL, KISTLER, BEACOM, \& HOPKINS}
\title{Revealing the High-Redshift Star Formation Rate with Gamma-Ray Bursts}
\author{Hasan Y{\"u}ksel\altaffilmark{1,2},
Matthew D. Kistler\altaffilmark{1,2},
John F. Beacom\altaffilmark{1,2,3},
and Andrew M. Hopkins\altaffilmark{4}}

\altaffiltext{1}{Dept.\ of Physics, The Ohio State University, 191 W.\ Woodruff Ave., Columbus, OH 43210}
\altaffiltext{2}{Center for Cosmology and Astro-Particle Physics, The Ohio State University, 191 W.\ Woodruff Ave., Columbus, OH 43210}
\altaffiltext{3}{Dept.\ of Astronomy, The Ohio State University, 140 W.\ 18th Ave., Columbus, OH 43210}
\altaffiltext{4}{School of Physics, The University of Sydney, NSW 2006, Australia}

%\email{kistler@mps.ohio-state.edu}
%\email{yuksel@mps.ohio-state.edu}
%\email{beacom@mps.ohio-state.edu}
%\email{ahopkins@physics.usyd.edu.au}
%\date{April 24, 2008}

\begin{abstract}
While the high-$z$ frontier of star formation rate (SFR) studies has advanced rapidly, direct measurements beyond $z \sim 4$ remain difficult, as shown by significant disagreements among different results.  Gamma-ray bursts, owing to their brightness and association with massive stars, offer hope of clarifying this situation, provided that the GRB rate can be properly related to the SFR.  The \textit{Swift} GRB data reveal an increasing evolution in the GRB rate relative to the SFR at intermediate $z$; taking this into account, we use the highest-$z$ GRB data to make a new determination of the SFR at $z = 4-7$.  Our results exceed the lowest direct SFR measurements, and imply that no steep drop exists in the SFR up to at least $z \sim 6.5$.  We discuss the implications of our result for cosmic reionization, the efficiency of the universe in producing stellar-mass black holes, and ``GRB feedback'' in star-forming hosts.
\end{abstract}


%%%%%%%%%%%%%%%%%%%%%%%%%%%%%%%%%%%%%%%%%%%%%%%%%%%%%%%%
%%%%%%%%%%%%%%%%%%%%%%%%%%%%%%%%%%%%%%%%%%%%%%%%%%%%%%%%

\section{Introduction}

\begin{figure}[t]
%\plotone{f1}
\centering\includegraphics[width=\linewidth,clip=true]{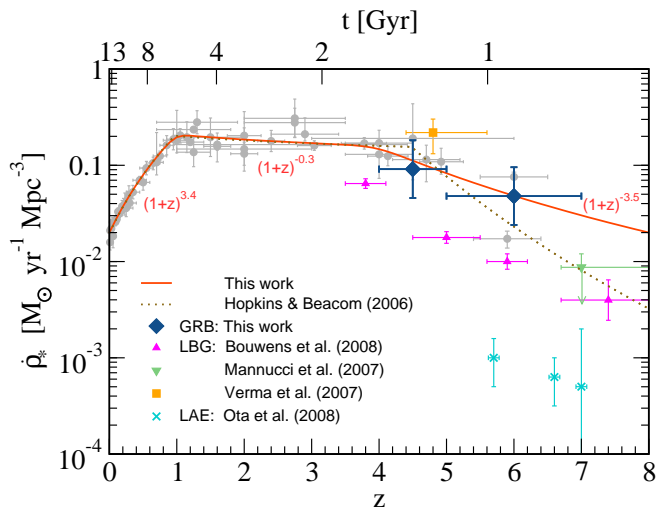}
\caption{The cosmic star formation history.  The compiled SFR data (light circles) and fit (dotted line) of \citet{Hopkins:2006bw} are shown, as well as newer high-$z$ data (the LAE points only sample Ly$\alpha$ Emitters).  The results of this work, as inferred using bright \textit{Swift} gamma-ray bursts, are shown with dark diamonds.  The solid line is our new high-$z$ fit given by Eq.~\ref{fit}.
\label{SFH}}
\end{figure}

The history of star formation in the universe is of intense interest to many in astrophysics, and it is natural to pursue pushing the boundary of observations to as early of times as possible.  Our understanding of this history is  increasing, with a consistent picture now emerging up to redshift $z \sim 4$, as summarized in Fig.~\ref{SFH}.  The cosmic star formation rate (SFR) measurements from the compilation of \citet{Hopkins:2006bw} are shown, along with new high-$z$ measurements based on observations of color-selected Lyman Break Galaxies (LBG) \citep{Bouwens08, Mannucci, Verma} and Ly$\alpha$ Emitters (LAE) \citep{Ota et al.(2008)}.  Much current interest is on this high-$z$ frontier, where the primeval stars that may be responsible for reionizing the universe reside.  Due to the difficulties of making and interpreting these measurements, different results disagree by more than their quoted uncertainties.

Instead of inferring the formation rate of massive stars from their observed populations, one may directly measure the SFR from their death rate, since their lives are short.  While it is not yet possible to detect ordinary core-collapse supernovae at high $z$, long-duration gamma-ray bursts, which have been shown to be associated with a special class of core-collapse supernovae \citep{Stanek:2003tw,Hjorth}, have been detected to $z = 6.3$.  The brightness of GRBs across a broad range of wavelengths makes them promising probes of the star formation history (SFH) (see, e.g., the early works of \citealt{Totani,Wijers:1998,Lamb,Blain:2000,Porciani,Bromm:2002}).  In the last few years, \textit{Swift}\footnote{See http://swift.gsfc.nasa.gov/docs/swift/archive/grb\_table.} \citep{Gehrels:2004am} has spearheaded the detection of GRBs over an unprecedentedly-wide redshift range, including many bursts at $z \gtrsim 4$.  Surprisingly, examination of the \textit{Swift} data reveals that GRB observations {\it are not} tracing the SFH directly, instead implying some kind of additional evolution \citep{Daigne, Le:2006pt, Yuksel:2006qb, Salvaterra, Guetta:2007, Kistler,  Salvaterra:2008sk}.

%%%%%%%%%%%%%%%%%%%%%%%%
%%%%%%%%%%%%%%%%%%%%%%%%%

GRBs can still reveal the overall amount of star formation, provided that we know how the GRB rate couples to the SFR.  In this study, we use the portion of the SFH that is sufficiently well-determined to probe the range beyond $z \simeq 4$.  We do this by relating the many bursts observed in $z \simeq 1-4$ to the corresponding SFR measurements, and by taking into account the possibility of additional evolution of the GRB rate relative to the SFR.  This calibration eliminates the need for prior knowledge of the absolute conversion factor between the SFR and the GRB rate and allows us to properly relate the GRB counts at $z \simeq 4-7$ to the SFR in that range.  Additionally, we make use of the estimated GRB luminosities to exclude faint low-$z$ GRBs that would not be visible in our high-$z$ sample, i.e., to compare ``apples to apples''.

Our results show that the SFR must be relatively high in the range $z = 4-7$ when compared to SFR measurements made using more conventional techniques.  While the GRB statistics at high $z$ are relatively low, they are high enough, permitting an approach complementary to other SFR determinations, which themselves must contend with presently unknown extinction corrections, cosmic variance, and selection effects, most importantly that flux-limited surveys necessarily probe the brightest galaxies, which may only contain a small fraction of the star formation activity at early epochs.  We discuss the implications these results in the context of reionization and broader applications.

%%%%%%%%%%%%%%%%%%%%%%%%%%%%%%%%%%%%%%%%%%%%%%%%%%%%%%%%
%%%%%%%%%%%%%%%%%%%%%%%%%%%%%%%%%%%%%%%%%%%%%%%%%%%%%%%%

\section{The GRB Technique}

The relationship between the comoving GRB and star formation rate densities can be parametrized as $\dot{n}_{\rm GRB}(z) = \mathcal{E}(z) \times \dot{\rho}_*(z)$, where $\mathcal{E}(z)$ reflects the fraction of stars that produce long-duration GRBs and any additional evolutionary effects.  Importantly, the SFH is well known for $z \lesssim 4$, and we use the \citet{Hopkins:2006bw} fit in this range.  The expected (all-sky) redshift distribution of GRBs can be cast as
\begin{equation}
	\frac{d\dot{N}}{dz} = F(z) 
	\frac{\mathcal{E}(z)\, \dot{\rho}_*(z)}{\left\langle f_{\rm beam}\right\rangle}
	\frac{dV/dz}{1+z}\,,
	\label{dnodz}
\end{equation}
where $0 < F(z) <1$ summarizes the ability both to detect the initial burst of gamma rays and to obtain a redshift from the optical afterglow, and includes observational factors such as instrumental sensitivities.  Beaming rendering some fraction of GRBs unobservable from Earth is accounted for through $\left\langle f_{\rm beam} \right\rangle$ \citep{Bloom,Firmani:2004fn,Kocevski}, the $1/(1+z)$ is due to cosmological time dilation of the observed rate, and $dV/dz$ is the comoving volume per unit redshift\footnote{$dV/dz = 4 \pi \, (c/H_0) \, d_c^2(z) / \sqrt{(1+z)^3\,\Omega_{\rm m} + \Omega_\Lambda}$, where $d_c$ is the comoving distance, $\Omega_{\rm m} = 0.3$, $\Omega_\Lambda = 0.7$, and $H_0 = 70$~km/s/Mpc.}.

Both $F(z)$ and $\mathcal{E}(z)$ are discussed in detail in \citet{Kistler}, in which it has been shown that the former can be set to a constant ($F_0$) by focusing on a bright subset of the GRB sample, allowing the latter to be parametrized as $\mathcal{E}(z) = \mathcal{E}_0 (1+z)^{\alpha}$, with $\alpha \simeq 1.5$ (an enhanced evolution of GRBs) preferred over $\alpha = 0$.  Here $\mathcal{E}_0$ is a (unknown) constant that includes the absolute conversion from the SFR to the GRB rate in a given GRB luminosity range.  The evolutionary trend described by $\alpha$ may arise from several mechanisms \citep{Kistler}, such as a GRB preference for low-metallicity environments \citep{Stanek:2006gc,Langer,Li:2007bq,Cen:2007xb}, and must be accounted for to characterize $\dot{n}_{\rm GRB}$.

%%%%%%%%%%%%%%%%%%%%%%%%%
\begin{figure}[t]
%\plotone{f2}
\centering\includegraphics[width=\linewidth,clip=true]{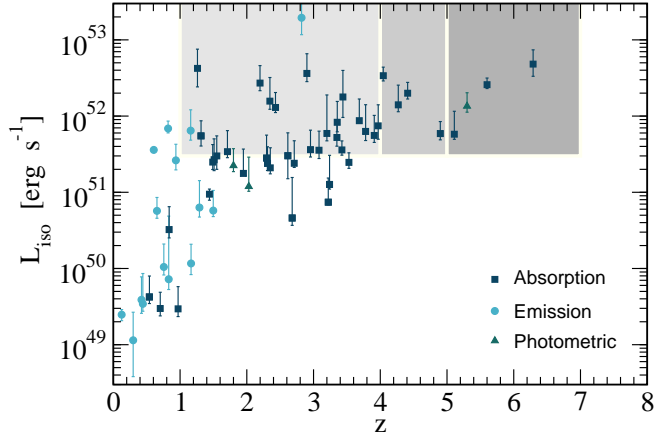}
\caption{The luminosity-redshift distribution of 63 \textit{Swift} GRBs (two $z < 0.1$ GRBs are not visible), as we determine from the data of Butler et al. (2007), marked according to the primary method of redshift determination.  The ratios of the number counts of GRBs in the three shaded boxes are used to estimate the SFR density at high $z$ (4--5, 5--7) by normalizing to the $z = 1-4$ SFR, as discussed in the text.  The boxes contain (21, 4, 4) bursts, respectively.
\label{Liso}}
\end{figure}
%%%%%%%%%%%%%%%%%%%%%%%%%

Fig.~\ref{Liso} shows the luminosity-redshift distribution of the 63 long-duration ($T_{90} > 2$~s) GRBs from the catalog of \citet{Butler:2007hw}.  We compute source-frame GRB luminosities, $L_{\rm iso} = E_{\rm iso} / [T_{90} / (1+z)]$, based on the values of $E_{\rm iso}$, the isotropic equivalent (uncorrected for beaming) $1-10^4$~keV energy release in the rest frame, and $T_{90}$, the interval containing 90\% of the prompt GRB emission.  Using another quantity, such as the peak isotropic equivalent flux (N.~Butler, private communication) yields quantitatively similar results.

As only very bright bursts can be seen from all redshifts, we define a cut, based on the luminosities of the high-$z$ events, of $L_{\rm iso} > 3 \times 10^{51}$ erg s$^{-1}$, for the subsample that we use to estimate the high-$z$ SFR\footnote{In \citet{Kistler}, where the focus was on intermediate $z$, a lower $L_{\rm iso}$ threshold was used.}.  (A naive normalization using all GRBs in $z = 0-4$ without regard to luminosity would yield values $\sim 2$ times smaller.)  This circumvents the need for an exact definition of the $z$-dependent threshold for detecting a GRB with a redshift measurement.  The complicated detection threshold of \textit{Swift} (e.g., \citealt{Band:2006fj}) and the human factor involved with optical observations makes a detailed treatment based on constructing the GRB luminosity function at low $z$ and marginalizing over this distribution to obtain the SFR at high-$z$ practically challenging, if not impossible.  The net effect would probably only be to make our eventual results somewhat larger.

The shaded boxes in Fig.~\ref{Liso} show three groups of GRBs defined by this $L_{\rm iso}$ cut in $z \!=\!$ $1-4$, $4-5$, and $5-7$ (the high-$z$ boxes chosen to have equal counts).  The GRBs in $z = 1-4$ act as a ``control group'' to base the GRB to SFR conversion, since this range has both good SFR measurements and GRB counting statistics.  We calculate the ``expected'' number of GRBs in this range as
\begin{eqnarray}
\mathcal{N}_{1-4}^{exp}
& = & \Delta t \frac{\Delta \Omega}{4\pi} \int_{1}^{4} dz\,  F(z) \, \mathcal{E}(z)  \frac{\dot{\rho}_*(z)}
{\left\langle f_{\rm beam}\right\rangle} \frac{dV/dz}{1+z} \nonumber \\
& = & \mathcal{A} \, \int_{1}^{4} dz\, \dot{\rho}_*(z)\, (1+z)^{\alpha} \, \frac{dV/dz}{1+z}\,,
\label{N1-4}
\end{eqnarray}
in which $\mathcal{A} = {\Delta t \, \Delta \Omega \, \mathcal{E}_0 \, F_0} / 4\pi {\left\langle f_{\rm beam} \right\rangle}$, an unknown normalization, depends on the total live-time, $\Delta t$, and the angular sky coverage, $\Delta \Omega$.  Using the average SFR density, $\left\langle \dot{\rho}_* \right\rangle_{z_1-z_2}$, a similar relation can be written for $z = 4-5$ and $5-7$ as
\begin{eqnarray}
\mathcal{N}_{z_1-z_2}^{exp}
& = &  \left\langle \dot{\rho}_* \right\rangle_{z_1-z_2} 
\mathcal{A} \, \int_{z_1}^{z_2} dz\, (1+z)^{\alpha} \, \frac{dV/dz}{1+z}\,.
\label{Nz1-z2}
\end{eqnarray}
Our interest is in $\left\langle \dot{\rho}_* \right\rangle_{z_1-z_2}$, which we find by dividing out $\mathcal{A}$, using Eq.~\ref{Nz1-z2}.  Taking the observed GRB counts, $\mathcal{N}_{z_1-z_2}^{obs}$, to be representative of the expected numbers, $\mathcal{N}_{z_1-z_2}^{exp}$, we find
\begin{equation}
\left\langle \dot{\rho}_* \right\rangle_{z_1-z_2} = 
\frac{\mathcal{N}_{z_1-z_2}^{obs}}{\mathcal{N}_{1-4}^{obs}} 
\frac{\int_{1}^{4} dz\, \frac{dV/dz}{1+z} \dot{\rho}_*(z)\, (1+z)^\alpha}{\int_{z_1}^{z_2} dz\,
\frac{dV/dz}{1+z} (1+z)^\alpha}\,,
\label{zratio}
\end{equation}
written in terms of the data for the groups of GRBs.  The decrease of $(dV/dz) / (1+z)$ at $z\gtrsim 1.5$ increasingly amplifies the significance of distant observed GRBs.

%%%%%%%%%%%%%%%%%%%%%%%%%%%%%%%%%%%%%%%%%%%%%%%%%%%%%%%%
%%%%%%%%%%%%%%%%%%%%%%%%%%%%%%%%%%%%%%%%%%%%%%%%%%%%%%%%

\section{The Inferred High-z Star Formation Rate}

We show our new determinations of the high-$z$ SFR in Fig.~\ref{SFH}, and how they compare with most of the 
conventional SFR results in this range\footnote{These measurements are based on the products of massive stars, and 
assume a \citet{Salpeter(1955)} stellar initial mass function.  A change in the low-mass IMF with $z$ would 
necessitate a rescaling (e.g., \citealt{Wilkins et al.(2008)}).}.  While evolution with $\alpha = 3$ (0) would 
decrease (increase) these values by a factor $\sim 2$, it would also disagree with $z<4$ data.  Other changes in the details of our analysis, while changing the results somewhat, do not change the important point that the high-$z$ SFR must be relatively large to accommodate the several observed high-$z$ GRBs.  These details include the exclusion of a particular GRB or changes in redshift assignments or ranges.  Taking into account the Poisson confidence interval for 4 observed events, and slightly enlarging it, we assign a statistical uncertainty of a factor 2 up or down.  As the number of GRBs increases, this uncertainty will correspondingly decrease, and more attention will need to be placed on systematics \footnote{While we use 8 z$>$4 GRBs from Butler et al. (2007), none were observed in the $\sim$ 1 year since then, indicative of a possible systematic effect.  This will not change our results more than the quoted uncertainties.}\footnote{Our normalizing to the whole range in z = 1--4 reduces the impact of a hypothesized deficit of GRBs in z = 1--2 \citep{Bloom,Coward:2007ra,Fiore:2007yk}}.  Our luminosity cut happens to restrict most of the GRB redshifts used to have been obtained by absorption lines, reducing one possible source of error.

We provide an update of the SFH fit of \citet{Hopkins:2006bw} based on our GRB results (solid line in Fig.~\ref{SFH}).  Here we adopt a continuous form of a broken power law,
\begin{eqnarray}
\dot{\rho}_*(z)
& = &  \dot\rho_0 \left[(1 + z)^{{a}{\eta} } + \left(\frac{1 + z}{B}\right)^{{b}{\eta}}
+ \left(\frac{1 + z}{C}\right)^{{c}{\eta} } \, \right]^{1/\eta}
\label{fit}
\end{eqnarray}
where $a = 3.4$ ($b = -0.3$, $c = -3.5$) is the logarithmic slope of the first (middle, last) piece, and the normalization is $\dot{\rho}_{0} = 0.02 \,M_\odot$~yr$^{-1}$~Mpc$^{-3}$.  Using $\eta \simeq -10$ smoothes the transitions, while $\eta \rightarrow -\infty$ would recover the kinkiness of the original form.  The breaks at $z_1= 1$ and $z_2 = 4$ correspond to $B = (1 + z_1)^{1-a/b} \simeq 5000$ and $C =(1 + z_1)^{(b-a)/c} (1 + z_2)^{1-b/c}\simeq 9$.

%%%%%%%%%%%%%%%%%%%%%%%%%%%%%%%%%%%%%%%%%%%%%%%%%%%%%%%%
%%%%%%%%%%%%%%%%%%%%%%%%%%%%%%%%%%%%%%%%%%%%%%%%%%%%%%%%

\section{Comparison to Other SFR Results}

Our SFR results, particularly the $z = 5-7$ point, are at the high end of SFR results obtained with more conventional methods, which show significant differences among themselves. Likely, these methods are measuring different populations of galaxies, with the GRBs probing low-luminosity galaxies that are otherwise not accounted for fully.  As detailed, the statistics of the GRB data are adequate, and it is unlikely that our GRB results are overestimated, and if anything, the true SFR is even larger.  We assume that a steep drop in the SFR is not being hidden by a steep rise in the fraction of stars that produce GRBs, beyond the evolution already taken into account; that would be even more interesting.

The extent of extinction by dust on high-$z$ SFR measurements is not yet strongly constrained.  There are a number of indications that dust is ubiquitous over the range $4 \lesssim z \lesssim 6$ (e.g., \citealt{Chary05, Ouchi, Ando}), and the dust correction in this range can be up to a factor of $\sim 2-3$.  At higher redshift, there are no strong observational constraints.  The dust corrections assumed in \citet{Bouwens07,Bouwens08} are generally small at high $z$; on the other hand, observations verify the existence of at least one heavily-obscured galaxy at $z \sim 6.6$ \citep{Chary05}, and UV-selected samples would be expected to be biased against such objects.  We also note that the recent mid-IR detection of the progenitor of SN 2008S implies that some short-lived stars do produce dust \citep{Prieto}.  For all of the LBG-based SFR results in Fig.~\ref{SFH}, we have included the dust corrections indicated by the respective authors.  While the true dust corrections may be larger, it seems likely that an additional factor of several is implausible.

The LBG surveys are most sensitive to the brightest galaxies, and great efforts are made to define the UV galaxy luminosity function (LF) and how well it is sampled at the faint end.  For example, the results shown in our Fig.~\ref{SFH} from \citet{Bouwens08} are integrated down $0.2 \, L_*$, for $L_*$ defined at $z = 3$.  For the three lowest-$z$ points, they also integrate down to $0.04 \, L_*$, yielding SFR results a factor of a few higher.  Correcting these measurements to account for the contributions from even fainter galaxies is necessary but difficult due to the poor observational constraints on the faint-end slope of the LF.  Depending on the value of this slope, and the lower limit of the integration, the necessary correction may be as little as a factor of two, or as much as an order of magnitude (see \citealt{Hopkins:2006}).  It also seems that Ly$\alpha$ Emitters do not contribute significantly to the total SFR.

%%%%%%%%%%%%%%%%%%%%%%%%%%%%%%%%%%%%%%%%%%%%%%%%%%%%%%%%
%%%%%%%%%%%%%%%%%%%%%%%%%%%%%%%%%%%%%%%%%%%%%%%%%%%%%%%%

\section{Discussion \& Conclusions}

We developed an empirical method for estimating the high-$z$ SFR using GRB counts, improving on earlier estimates of the high-$z$ SFR using GRB data (e.g., \citealt{Berger:2005, Natarajan:2005, Jakobsson, Le:2006pt, Chary}) in several ways, not least of which by using significantly updated SFR \citep{Hopkins:2006bw} and/or GRB \citep{Butler:2007hw} data.  Taking advantage of the improved knowledge of the SFR at intermediate $z$, we were able to move beyond the assumption of a simple one-to-one correspondence between the GRB rate and the SFR, accounting for an increasing evolutionary trend.  The higher statistics of the recent \textit{Swift} GRB data allowed the use of luminosity cuts to fairly compare GRBs in the full $z$ range, eliminating the uncertainty of the unknown GRB luminosity function.  By comparing the counts of GRBs at different $z$ ranges, normalized to SFR data at intermediate $z$, we based our results squarely on data, eliminating the need for knowledge of the absolute fraction of stars that produce GRBs.

Our GRB-based SFR value in $z = 4-5$ is comparable to more conventional results, which may be taken as a validation
of our method, i.e., possible selection effects have been adequately controlled (see \citealt{Kistler} for more details), while our SFR value in $z = 5-7$ is larger, which has important implications.  A significant population of low-luminosity star-forming galaxies picked out by GRBs but missed in other surveys could reconcile these differences (searches utilizing cluster lensing, e.g., \citealt{Richard:2008at}, may help, if cosmic variance can be understood).  \citet{Yan04} argue that the correction for the faint-end of the LF should be large to correct for dwarf galaxies missed in LBG surveys.  Since GRBs are observed to favor subluminous host galaxies \citep{Fynbo:2003sx, Le Floc'h:2003yp, Fruchter}, they may be probing such faint galaxies.

The universe was fully reionized by $z \simeq 6$ \citep{Wyithe et al.(2005), Fan et al.(2006)}, and it appears that AGN could not have been responsible \citep{Hopkins et al.(2008)}.  Can stars have reionized the universe?  In \citet{Madau et al.(1999)}, the SFR density required to produce a sufficient ionizing photon flux is parametrized by two factors: the photon escape fraction ($f_{\rm esc}$) and the clumpiness of the IGM ($\mathcal{C}$).  These enter as a ratio, $\mathcal{C}/f_{\rm esc}$, which is not known precisely.  For $\mathcal{C}/f_{\rm esc} \lesssim 30$, the required SFR at $z=6$ is $\dot{\rho}_*(z = 6) \gtrsim 0.03\, M_\odot$~yr$^{-1}$~Mpc$^{-3}$, just at the level of our GRB-based SFR.

This ratio may be higher (due to either $\mathcal{C}$ or $f_{\rm esc}$), however, which would render our measured value of $\dot{\rho}_*(z\!=\!6)$ too low, reintroducing the concern that star formation may be insufficient to achieve reionization (e.g., \citealt{Gnedin(2008)}, C.~A.~Faucher-Giguere et al. 2008, in preparation).  Is there any way around this?  One can plausibly increase our SFR determination by considering the increasing incompleteness of the GRB sample with $z$ due to our $L_{\rm iso}$ cut, since for fixed luminosity, higher-$z$ bursts are relatively more difficult to detect \citep{Butler:2007hw} and the universe will at some $z$ quickly become opaque blueward of Ly$\alpha$ (e.g., \citealt{Ciardi:2000by}).  If there is a {\it maximum} metallicity for forming a GRB, then the evolutionary trend may saturate at high enough redshift; arbitrarily assuming no evolution beyond $z \gtrsim 4$ would give an additional factor of $\lesssim 2$.  Finally, requiring a {\it minimum} amount of metals in a star for a successful GRB would suppress the GRB rate at increasingly high-$z$, necessitating a higher underlying SFR.  A similar argument could be made concerning whether early, metal-poor stars possessed sufficient angular momentum.  Overall, more data are needed concerning the varying ionization fraction with redshift to establish if star formation is quantitatively acceptable as an explanation for reionization, instead of simply by default if AGN have been eliminated.

The possibility of continuing evolution in the GRB rate relative to the SFR invites some astrophysical speculations.  While the overall black hole production rate in core-collapse supernovae is poorly understood empirically \citep{Kochanek et al.(2008)}, bright GRBs might be regarded as tracing the large angular momentum end of the black hole birth distribution (e.g., for collapsars; \citealt{MacFadyen}).  Evolution would mean that the high-$z$ universe was more efficient at producing such black holes than usually considered.  This may have implications for the nucleosynthetic yields from these explosions (e.g., \citealt{nomoto}).  The effect of supernovae on the gas in high-$z$ galaxies is an ongoing field of research (e.g., \citealt{Whalen et al.(2008)}).  Following from the considerations above, it may be that small galaxies had a disproportionately large rate of GRBs relative to normal supernovae.  In this case, it would be interesting to examine the fate of such galaxies after including multiple injections of highly-asymmetric, relativistic ejecta, as well as implications for enrichment of the IGM (i.e., ``GRB feedback'').

%%%%%%%%%%%%%%%%%%%%%%%%%%%%%%%%%%%%%%%%%%%%%%%%%%%%%%%%
%%%%%%%%%%%%%%%%%%%%%%%%%%%%%%%%%%%%%%%%%%%%%%%%%%%%%%%%

\acknowledgments
We thank Nat Butler, Kris Stanek, Todd Thompson, and Haojing Yan for helpful discussions.
We acknowledge use of the \textit{Swift} public archive.
HY and JFB were supported by NSF CAREER Grant PHY-0547102 to JFB, MDK by DOE Grant DE-FG02-91ER40690, and AMH by the Australian Research Council in the form of a QEII Fellowship (DP0557850).

%%%%%%%%%%%%%%%%%%%%%%%%%%%%%%%%%%%%%%%%%%%%%%%%%%%%%%%%
%%%%%%%%%%%%%%%%%%%%%%%%%%%%%%%%%%%%%%%%%%%%%%%%%%%%%%%%

%%%%%%%%%%%%%%%%%%%%%%%%%%%%%%%%%%%%%%%%%%%%%%%%%%%%%%%%
%%%%%%%%%%%%%%%%%%%%%%%%%%%%%%%%%%%%%%%%%%%%%%%%%%%%%%%%
%\clearpage

\end{document}